\documentclass[amstex,thmsa,11pt,sw20bams]{article}%
\usepackage{amsmath}
\usepackage{amsfonts}
\usepackage[utf8]{inputenc}
\usepackage{graphicx}
\usepackage[francais]{babel}
\usepackage{amssymb}%
\providecommand{\U}[1]{\protect\rule{.1in}{.1in}}
\ifx\pdfoutput\relax\let\pdfoutput=\undefined\fi
\newcount\msipdfoutput
\ifx\pdfoutput\undefined\else
\ifcase\pdfoutput\else
\ifx\paperwidth\undefined\else
\ifdim\paperheight=0pt\relax\else\pdfpageheight\paperheight\fi
\ifdim\paperwidth=0pt\relax\else\pdfpagewidth\paperwidth\fi
\fi\fi\fi
\begin{document}

\title{Binary Trees and Taxicab Correspondence Analysis of Extremely Sparse
Binary Textual Data: A Case Study }
\author{Choulakian V, Universit\'{e} de Moncton,
vartan.choulakian@umoncton.ca \and Allard J, Universit\'{e} de Moncton,
jacques.allard@gmail.com \and Kenett R, Samuel Neaman Institute, Technion,
ron@kpa-group.com}
\maketitle

\begin{abstract}
This is a case study, where Taxicab Correspondence Analysis reveals that the
underlying structure of an extremely sparse binary textual data set can be
represented by a binary tree, where the nodes representing clusters of words
can be interpreted as topics. The textual data set represents Israel's
Declaration of Independence text and 40 diverse Israeli Interviewees. The
analysis provides for a compare and contrast study of textual data coming
from two different sources. Furthermore, we propose an adjusted sparsity
index which takes into account the size of the data table.

\end{abstract}

\section{Introduction}

The motivation of this paper is the statistical analysis of an extremely
sparse binary (presence/absence) 0/1 valued textual data set $\mathbf{Z}%
=(z_{ij})$ of size $1730\times 343;$ where $z_{ij}=1$ means the presence of
the token (word) or term (collection of words) $i$ in the document (phrase) $%
j,$ and $z_{ij}=0$ means the absence of the token (word) or term (collection
of words) $i$ in the document (phrase) $j.$ The data set is constructed from
(40+1) texts: the 40 texts represent interviews done in Israel during the
Covid in 2021, plus the text of the Declaration of Independence (DOI) of the
State of Israel announced by David Ben Gurion in 1948. It provides an
approach to compare and contrast unstructured data from different sources.
This ability to integrate the analysis of text from different origins is
demonstrated here with a case study.

According to Kenett et al. (2023) "40 personal in-depth interviews with a
series of well-known and influential Israeli individuals from various groups
and genders, who are involved in various areas of endeavour, positioned all
over the political spectrum and are of diverse ages and worldviews: a
`representative sample' (that includes) male and female, Jewish, Druze, and
Arab, secular, religious and ultra-Orthodox, and young and older respondents
holding a broad range of social and political positions. Those interviewed
included senior academics, writers and scientists, former politicians, media
figures and publicists, rabbis, economists, social entrepreneurs,
strategists, and retired military officers. Most of the interviews were
conducted in January-February 2021 via Zoom, and typically lasted between
sixty to ninety minutes. They expressed their viewpoints and attitudes
towards the question representing `organizing principles' and values on
which Israeli society should be predicated in the decades to come. A
qualitative, intuitive, analysis of the interviews led the researchers to
the conclusion that there is indeed a broad common denominator among those
with a wide variety of opinions, which is closely linked with the major
principles of the Declaration of Independence (DOI) of the State of Israel."

The data has been analyzed by Kenett et al. (2023) by latent semantic
analysis (LSA), which consists of singular value decomposition (SVD) of the
tf-idf coded data set.

The approach differs in this paper: It is based on the joint use of
correspondence analysis (CA) and its robust $l_{1}$ variant named taxicab CA
(TCA). TCA is able to reveal the underlying structure of the extremely
sparse data in a complete binary tree; which represents the interpretable
diagonal blocks structure of the data set; that is, biclustring of the rows
and the columns simultaneously. It is important to note that visualization
via maps plays pivotal role in the interpretation by CA and TCA approaches.
Furthermore, we propose an adjusted sparsity index which takes into account
the size of the data table.

In section 2, we analyze in detail the Declaration of Independence (DOI)
text by CA and TCA, where all the details of the computation are shown. In
sections 3, 4\ and 5, we analyze the full (40+1) corpus of texts from
interviews; in section 6 we propose the adjusted sparsity index; and we
conclude in section 7.

Benz\'{e}cri (1966) is the reference source on the linguistic foundations of
CA; while the reference book on applied textual data analysis by CA is Benz%
\'{e}cri (1981), where binary textual data are discussed in pages 259-359.
Lebart (2024) describes the analysis of binary data by PCA and CA: With the
influence of rare observations (words) we often see the common practice of
eliminating words with marginal frequencies less than 10. See also Qi et al.
(2023, page 17). This practice in CA of textual data dates since Benz\'{e}%
cri (1981); where the elimination of rare columns or rows is done to obtain
interpretable maps by CA. TCA, being robust, eliminates only rows or columns
with zero marginals; see Choulakian et al. (2006, 2023) and this paper.

\section{The analysis of the Declaration of Independence}

\textquotedblright The Declaration of Independence (DOI) has, basically,
three parts :

Part a) Historical: It links the State of Israel to its historical-Biblical
roots;

Part b) Operational: It includes statements about how the state would
operate;

Part c) Visionary: This is the most pivotal section, comprising statements
protecting civil liberties and expressing aspirations for peace and
coexistence.

In the next subsection, we reproduce DOI text from Kenett et al. (2023): It
is composed of 21 rows (phrases) and 43 different words (terms) will
represent the columns of the binary table to be analyzed in detail by CA and
TCA.

\subsection{Declaration of Independence text}

R1) THE STATE OF ISRAEL will be open for Jewish immigration and for the
Ingathering of the Exiles

R2) it will foster the development of the country for the benefit of all its
inhabitants;

R3) it will be based on freedom, justice and peace as envisaged by the
prophets of Israel;

R4) it will ensure complete equality of social and political rights to all
its inhabitants irrespective of religion, race or sex;

R5) it will guarantee freedom of religion, conscience, language, education
and culture;

*R6) it will safeguard the Holy Places of all religions; and

*R7) it will be faithful to the principles of the Charter of the United
Nations.

R8) THE STATE OF ISRAEL is prepared to cooperate with the agencies and
representatives of the United Nations in implementing the resolution of the
General Assembly of the 29th November, 1947, and

R9) will take steps to bring about the economic union of the whole of
Eretz-Israel.

R10) WE APPEAL to the United Nations to assist the Jewish people in the
building-up of its State and

R11) to receive the State of Israel into the comity of nations.

R12) WE APPEAL - in the very midst of the onslaught launched against us now
for months --

R13) to the Arab inhabitants of the State of Israel to preserve peace and
participate in the upbuilding of the State

R14) on the basis of full and equal citizenship and due representation in
all its provisional and permanent institutions.

R15) WE EXTEND our hand to all neighboring states and their peoples in an
offer of peace and good neighborliness, and

R16) appeal to them to establish bonds of cooperation and mutual help with
the sovereign Jewish people settled in its own land.

R17) The State of Israel is prepared to do its share in a common effort for
the advancement of the entire Middle East.

R18) WE APPEAL to the Jewish people throughout the Diaspora to rally round
the Jews of Eretz-Israel

*R19) in the tasks of immigration and upbuilding and to stand by them

R20) in the great struggle for the realization of the age-old dream - the
redemption of Israel.

R21) HEREBY DECLARE THE ESTABLISHMENT OF A JEWISH STATE IN ERETZ-ISRAEL, TO
BE KNOWN AS THE STATE OF ISRAEL.

\subsection{Construction of the binary data table}

The DOI text is divided into 21 phrases numbered from R1 to R21. 18 phrases
are described by 43 apriori chosen words, from which the presence/absence
(0/1) coded table $\mathbf{Z}$ of size $I\times J=18\times 43$ is formed.
Such a table is called a document term matrix (DTM) with binary weights.

Phrases R$6,\ $R$7\ $and$\ $R$19$, starred above, are deleted and not
included in the statistical analysis since they do not include any of the 43
words analyzed here.

By applying Benz\'{e}cri's principle of distributional equivalence, $\mathbf{%
Z}$ is equivalent to the contingency table $\mathbf{N}$ of size $I\times
J=18\times 24.$ The apparent sparsity of $\mathbf{Z}$ is 92.38\%; while the
sparsity of $\mathbf{N}$ is 90.74\%, which represents the sparsity of the
data. The contingency table $\mathbf{N}$ can be found in the Appendix.

Table 1 represents the distributions of the row and column marginals of the
CA (TCA) equivalent data sets $\mathbf{Z}$ and $\mathbf{N}$.

\begin{tabular}{c|ccccc|cccccc}
\multicolumn{12}{c}{\textbf{Table 1: CA equivalent data sets }$\mathbf{Z}$%
\textbf{\ and }$\mathbf{N}$ \textbf{of DOI}.} \\ \hline
& \multicolumn{11}{|c}{column marginals} \\ \hline
& \multicolumn{5}{|c|}{$\mathbf{Z}$ of size $18\times 43$} & 
\multicolumn{6}{|c}{$\mathbf{N}$ of size $18\times 24$} \\ \hline
values & 1 & 2 & 3 & 5 & 6 & $1$ & $2$ & $3$ & $4$ & $5$ & $6$ \\ 
counts & 36 & 3 & 2 & 1 & 1 & 4 & 12 & 4 & 2 & 1 & 1 \\ \hline
& \multicolumn{5}{|c|}{row marginals of $\mathbf{Z}$ or $\mathbf{N}$} & 
\multicolumn{6}{|c}{} \\ \cline{1-6}
values & 1 & 2 & 3 & 4 & 5 &  &  &  &  &  &  \\ 
counts & $1$ & 3 & 6 & 6 & 2 &  &  &  &  &  &  \\ \cline{1-6}
\end{tabular}

A hapax is a word used once in the text; that is its marginal count is 1. It
is a common practice in CA, specially in textual data analysis by CA, to
eliminate hapaxes; see for instance Lebart (2024). If we eliminate hapaxes
in $\mathbf{Z}$, then we eliminate 36 columns (words), see Table 1. If we
eliminate hapaxes in $\mathbf{N}$, then we eliminate 4\ columns (words). For
instance, the 4\ hapaxes (social, political, equality, rights) in \textbf{Z}
are no more hapaxes in $\mathbf{N}$: they represent the term (social +
political + equality + rights) and are found in phrase R4; moreover they
represent important concept of human and democratic rights; thus eliminating
them carries information loss. Similarly for the term (economic + take +
bring + whole) found in the phrase R9.

\subsection{Correspondence analysis (CA)}

We use the R package ca of Greenacre et al. (2022) to analyze the contents
of \textbf{N}. First, we look at the singular values:

 res.caNmerged\$sv

[1] 1.0000000 1.0000000 1.0000000 0.9806344 0.9481851 0.9327178 0.9292113
0.8599820

[9] 0.8458454 0.8164966 0.7975632 0.7595974 0.7506988 0.7148211 0.6614619
0.5756657

The first 3 singular values are equal to 1, which shows that the underlying
structure of $\mathbf{N}$ is 4-blocks diagonal (see the CA map in the
appendix): 
\begin{equation*}
\mathbf{N=}\left( 
\begin{array}{cccc}
\mathbf{B}_{1} &  &  &  \\ 
& \mathbf{B}_{2} &  &  \\ 
&  & \mathbf{B}_{3} &  \\ 
&  &  & \mathbf{B}_{4}%
\end{array}%
\right) ,
\end{equation*}%
where each of the 3 blocks $\mathbf{B}_{1},\mathbf{B}_{2}$ and $\mathbf{B}%
_{3}$ is made of the cells (R2,C11), (R12,C9), (R14,C20), where

C11 is the term $"country+development"$ in phrase R2

C9 is the term $"us+now"$ in phrase R12

C20 is the term $"basis+full+equal"$ in phrase R14;

and the remaining cells form the 4th block, $\mathbf{B}_{4}$ of size $%
15\times 21.$ So CA of \textbf{N} is equivalent to CA of $\mathbf{B}_{4}.$ $%
\mathbf{B}_{4}$ is of size $15\times 21$ and its sparsity is 88.25\%.

The singular values resulting from CA of $\mathbf{B}_{4}$ are identical to
to the singular values of CA of $\mathbf{N}$ which are different from 1, as
can be seen below:

res.caB4\$sv

[1] 0.9806344 0.9481851 0.9327178 0.9292113 0.8599820 0.8458454 0.8164966
0.7975632

[9] 0.7595974 0.7506988 0.7148211 0.6614619 0.5756657 0.3178244

Figure 1 represents the principal map of dimensions 1 and 2 obtained from CA
of $\mathbf{B}_{4},$ which is identical to the principal map of dimensions 4
and 5 of CA of $\mathbf{N}$.

\begin{figure}[h]
\label{fig:MagEngT}
\centering 
\includegraphics[scale=0.5]{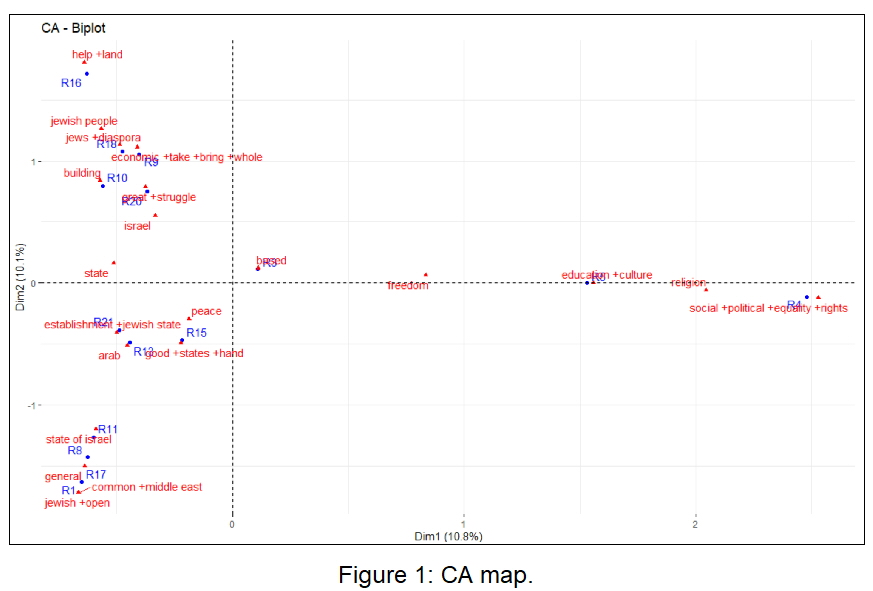}  
\end{figure}

Visually, we see that there are
at least 3 clusters in Figure 1. We show that there are 4 clusters: The one
on the right is evident, while the boundaries of the 3 clusters on the left
are fuzzy. A clearer picture of the 4\ clusters is obtained by TCA, which we
present in the next subsection.

\subsection{Taxicab correspondence analysis (TCA)}

We use the R package TaxicabCA of Allard and Choulakian (2019) to obtain TCA
Figure 2, which displays TCA principal map of \textbf{B}$_{4},$ and Figure 3
displays the associated TCA contribution principal map of \textbf{B}$_{4}.$
It is evident that there are 4 clusters, where one cluster is found in each
quadrant. The 4 extreme points in Figure 3\ succinctly describe the topics
elaborated in the DOI: \textbf{State of Israel} (political), \textbf{jewish
people} (ethnic identity), \textbf{economic} (development) and \textbf{%
social+political+equality+rights} (human rights). Here, we describe in more
detail the 4 clusters (topics). The interpretation of each cluster is based
on the contribution (displayed in parenthesis) of each point (row or
column). Note that the 4 clusters are denoted by \textbf{PP, PN, NP} and 
\textbf{NN}, where $N=negative$ and $P=positive$.\bigskip\ 

\textbf{Cluster NP in the upper left quadrant }: The words \{\textbf{social
+political +equality +rights} (7.45), \textbf{education +culture} (3.36), 
\textbf{religion} (2.92), \textbf{freedom} (2.56)\} appear in the two
phrases \textbf{R4} (5.86), \textbf{R5} (4.51)

R4) it will ensure complete \textbf{equality} of \textbf{social} and \textbf{%
political} \textbf{rights} to all its inhabitants irrespective of \textbf{%
religion}, race or sex

R5) it will guarantee \textbf{freedom} of \textbf{religion}, conscience,
language, \textbf{education} and \textbf{culture}\bigskip\ 

\textbf{Cluster PP in the upper right quadrant }: The words \{\textbf{state
of israel} (4.46), \textbf{establishment +jewish state} (3.05), \textbf{%
Jewish+open} (3.05), \textbf{common + middle east} (3.06), \textbf{general}
(0.72)\} appear in the phrases \textbf{R21} (2.93), \textbf{R1} (2.87), 
\textbf{R17} (1.71), \textbf{R8} (1.14), \textbf{R11} (0.57):

R21) HEREBY DECLARE \textbf{THE ESTABLISHMENT OF A JEWISH STATE} IN
ERETZ-ISRAEL, TO BE KNOWN AS THE \textbf{STATE OF ISRAEL}

R1) THE \textbf{STATE OF ISRAEL} will be \textbf{open} for \textbf{Jewish}
immigration and for the Ingathering of the Exiles

R17) The \textbf{State of Israel} is prepared to do its share in a \textbf{%
common} effort for the advancement of the entire \textbf{Middle East}

R8) THE \textbf{STATE OF ISRAEL} is prepared to cooperate with the agencies
and representatives of the United Nations in implementing the resolution of
the \textbf{General} Assembly of the 29th November, 1947

R11) to receive the \textbf{State of Israel} into the comity of
nations\bigskip\ 

\textbf{Cluster NN in the lower left quadrant }: The words \{\textbf{%
economic + take + bring+whole} (5.65), \textbf{good +states +hand} (5.10), 
\textbf{peace} (2.97), \textbf{great +struggle} (2.82), \textbf{israel}
(2.81), \textbf{based} (1.43)\} appear in in the \textbf{R9} (5.28), \textbf{%
R15} (4.14), \textbf{R20} (3.17), \textbf{R3} (2.74) phrases:

R9) will \textbf{take} steps to \textbf{bring} about the \textbf{economic}
union of the \textbf{whole} of Eretz-Israel

R15) WE EXTEND our \textbf{hand} to all neighboring \textbf{states} and
their peoples in an offer of \textbf{peace} and \textbf{good}
neighborliness, and

R20) in the \textbf{great struggle} for the realization of the age-old dream
- the redemption of \textbf{Israel}

R3) it will be \textbf{based} on freedom, justice and peace as envisaged by
the prophets of Israel\bigskip\ 

\textbf{Cluster PN in the lower right quadrant }: The words \{\textbf{jewish
people} (4.01), \textbf{help +land} (3.21), \textbf{jews +diaspora }(3.21), 
\textbf{state} (3.05), \textbf{building} (1.53), \textbf{arab} (1.53)\}
appear in the phrases \textbf{R16} (3.50), \textbf{R18} (3.36), \textbf{R13}
(2.98), \textbf{R10} (2.96):

R16) appeal to them to establish bonds of cooperation and mutual \textbf{help%
} with the sovereign \textbf{Jewish people} settled in its own \textbf{land}

R18) WE APPEAL to the \textbf{Jewish people} throughout the \textbf{Diaspora}
to rally round the \textbf{Jews} of Eretz-Israel

R13) to the \textbf{Arab} inhabitants of the \textbf{State} of Israel to
preserve peace and participate in the up\textbf{building} of the \textbf{%
State}

R10) WE APPEAL to the United Nations to assist the \textbf{Jewish people} in
the \textbf{building}-up of its \textbf{State} and

\bigskip

\begin{figure}[h]
\label{fig:MagEngT}
\centering 
\includegraphics[scale=0.5]{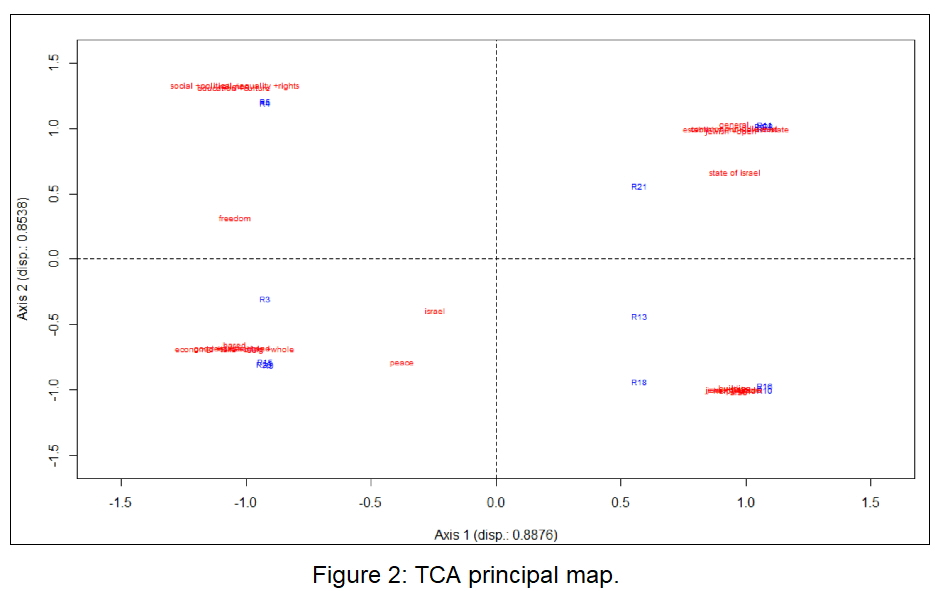}  
\end{figure}

\begin{figure}[h]
\label{fig:MagEngT}
\centering 
\includegraphics[scale=0.5]{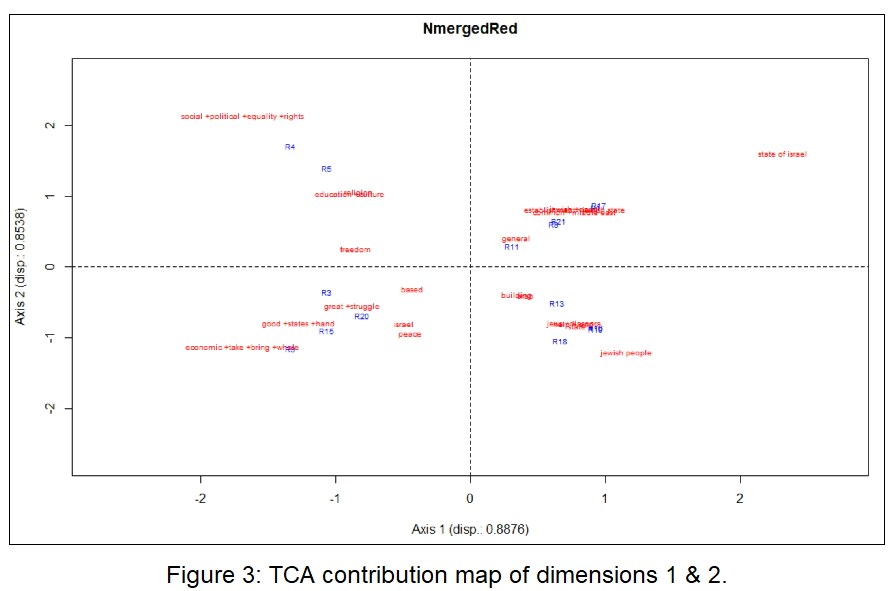}  
\end{figure}

\subsection{Underlying structure of \textbf{B}$_{4}$}

According to Benz\'{e}cri (1973), there is a \textit{lateral ordering} in
the structure of the data set $\mathbf{B}_{4}$ because the value of the
first principal singular value $\rho _{1}=0.9806>0.837$. This lateral
ordering can be of 2\ kinds: \textit{quasi-multi blocks diagonal} or \textit{%
quasi-diagonal (Guttman effect)}. For extremely sparse data sets, such as of 
$\mathbf{B}_{4}$ whose sparsity is 88.25\%, it is often difficult to
distinguish between these 2 cases.

\begin{tabular}{lllll}
\multicolumn{5}{l}{\textbf{Table 2: QSR (\%) for the first 4 dimensions of
DOI.}} \\ \hline\hline
&  & TCA &  &  \\ 
Axis $\alpha $ & $QSR_{\alpha }(V_{+}U_{+},V_{-}U_{-})$ & $QSR_{\alpha
}(V_{-}U_{+},V_{+}U_{-})$ & $QSR_{\alpha }$ & $\delta _{\alpha }$ \\ \hline
1 & (39.5,\textbf{\ }36.9) & (\textbf{-100},\ -74.3\textbf{)} & \textbf{52.7}
& \textbf{0.888} \\ 
2 & (51.1,\ 33.6\textbf{)} & (-56.2\textbf{,\ }-74.5) & \textbf{49.6} & 
\textbf{0.854} \\ 
3 & (31.6,\ 35.1) & (-73.5, -87.1) & 46.9 & 0.836 \\ 
4 & (46.0, 45.2) & (-58.0, -58.4) & 51.1 & 0.802 \\ \hline
\end{tabular}

In the framework of TCA, the QSR index also can be helpful in deriving the
underlying structure of the extremely sparse set $\mathbf{B}_{4}.$ Table 2
displays the QSR indices in \%. For the first principal dimension, we note
that $QSR_{1}(V_{-}U_{+})=-100\%$; this means that the underlying structure
of $\mathbf{B}_{4}$ is reducible to:

$\mathbf{B}_{4}\equiv \left( 
\begin{array}{cc}
\mathbf{A} & \mathbf{0} \\ 
\mathbf{C} & \mathbf{D}%
\end{array}%
\right) .$

\subsection{Binary Tree Representation of the 4 clusters}

Principal dimensions define the binary tree in any dimension reduction
approach, in particular in TCA of extremely sparse data set.

Based on the first principal dimension, we represent the matrix $\mathbf{B}%
_{4}$ as a binary tree with 2 nodes

$\mathbf{B}_{4}\equiv \left( 
\begin{array}{cc}
\mathbf{N} &  \\ 
& \mathbf{P}%
\end{array}%
\right) ,$

where \textbf{N} represents the subtable representing the rows and the
columns on the left side in Figure 2, and \textbf{P} represents the subtable
representing the rows and the columns on the right side in Figure 2.

Based on the second principal dimension, we represent the matrix $\mathbf{B}%
_{4}$ as a binary tree with 4 nodes

$\mathbf{B}_{4}\equiv \left( 
\begin{array}{cccc}
\mathbf{NN} &  &  &  \\ 
& \mathbf{NP} &  &  \\ 
&  & \mathbf{PN} &  \\ 
&  &  & \mathbf{PP}%
\end{array}%
\right) .$

\section{The analysis of the binary data of (40+1) texts}

We follow exactly the same process in the analysis-visualization of the
corpus of (40+1) DTM data; where 40 represents the number of diverse Israeli
interviewees and 1 the DOI text.

\subsection{General statistics}

The original presence/absence (0/1) data set of size $1730\times 343,$ where
1730 represents the number of phrases in (40+1) texts and 343 the number of
distinct words (or terms). It has 70 rows with 0 marginal counts; by
deleting them, we obtain the (0/1) matrix \textbf{Z} of size $1660\times 343$%
, whose apparent sparsity index is $98.72\%$. By Benz\'{e}cri's principle of
distributional equivalence \textbf{Z} is equivalent to the contingency table 
\textbf{N} of size $1605\times 343$ with the sparsity index of $98.68\%$.
Table 3 represents a comparison of the row and column marginals of \textbf{Z}
and \textbf{N}; where we notice that the number of hapaxes in \textbf{Z} is
169, slightly more than twice 83, the number of hapaxes in \textbf{N}.

\begin{tabular}{c|ccccccccccc}
\multicolumn{12}{c}{\textbf{Table 3: CA equivalent data sets }$\mathbf{Z}$%
\textbf{\ and }$\mathbf{N}$ \textbf{of (40+1) texts}.} \\ \hline
& \multicolumn{11}{|c}{row marginals of $\mathbf{Z}$ of size $1660\times 343$%
} \\ \hline
values & 1 & 2 & 3 & 4 & 5 & $6$ & $7$ & \multicolumn{4}{c}{$[9:17]$} \\ 
counts & 169 & 238 & 303 & 268 & 239 & 133 & 114 & \multicolumn{4}{c}{196}
\\ \hline
& \multicolumn{11}{|c}{row marginals of $\mathbf{N}$ of size $1605\times 343$%
} \\ \hline
values & 1 & 2 & 3 & 4 & 5 & \multicolumn{1}{|c}{$6$} & $7$ & 
\multicolumn{4}{c}{$[9:17]$} \\ 
counts & 83 & 254 & 311 & 274 & 239 & \multicolumn{1}{|c}{134} & 114 & 
\multicolumn{4}{c}{196} \\ \hline
\multicolumn{12}{c}{summary of column marginals of $\mathbf{Z}$ or $\mathbf{N%
}$} \\ \cline{1-6}
& Min & Q$_{1}$ & Median & Mean & Q$_{3}$ & Max & \multicolumn{1}{|c}{} &  & 
&  &  \\ 
values & $7$ & $12$ & $15$ & $21.31$ & $24$ & $143$ & \multicolumn{1}{|c}{}
&  &  &  &  \\ \cline{1-7}
\end{tabular}

\begin{figure}[h]
\label{fig:MagEngT}
\centering 
\includegraphics[scale=0.5]{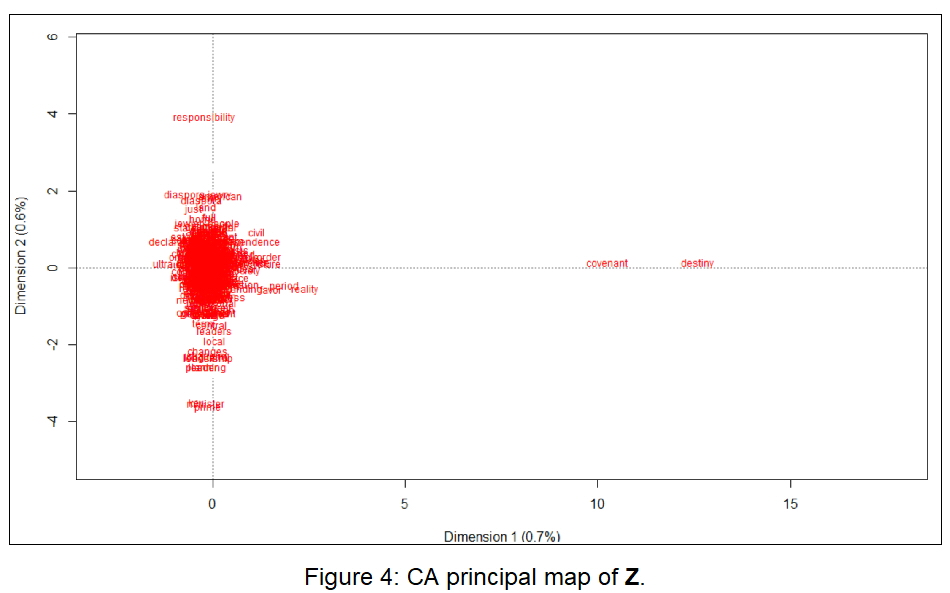}  
\end{figure}

\subsection{Correspondence Analysis of \textbf{N or Z}}

Figure 4 displays the principal map of the 343 columns (words) obtained from
CA of \textbf{N }or\textbf{\ Z}, which is difficult to interpret. However,
by checking in Table 4 the singular values $\rho $, the correlation
coefficients of the row and column factor scores, we note that they are
quite high, the first few around 0.7; which shows that there is some
underlying structure which appears by using TCA.\bigskip

\begin{tabular}{lllll}
\multicolumn{5}{l}{\textbf{Table 4: CA dispersion values of the first 4
principal axes.}} \\ \hline\hline
&  & Axis &  &  \\ 
data set & $\rho _{1}$ & $\rho _{2}$ & $\rho _{3}$ & $\rho _{4}$ \\ \hline
\textbf{Z} or \textbf{N} & 0.731 & 0.693 & 0.676 & 0.674 \\ \hline
\textbf{DataPosPos} & 0.812 & 0.797 & 0.790 & 0.786 \\ \hline
\textbf{DataPosNeg} & 0.907 & 0.902 & 0.870 & 0.836 \\ \hline
\textbf{DataNegNeg} & 0.866 & 0.848 & 0.842 & 0.830 \\ \hline
\textbf{DataNegPos} & 0.862 & 0.846 & 0.836 & 0.832 \\ \hline
\end{tabular}

\begin{figure}[h]
\label{fig:MagEngT}
\centering 
\includegraphics[scale=0.5]{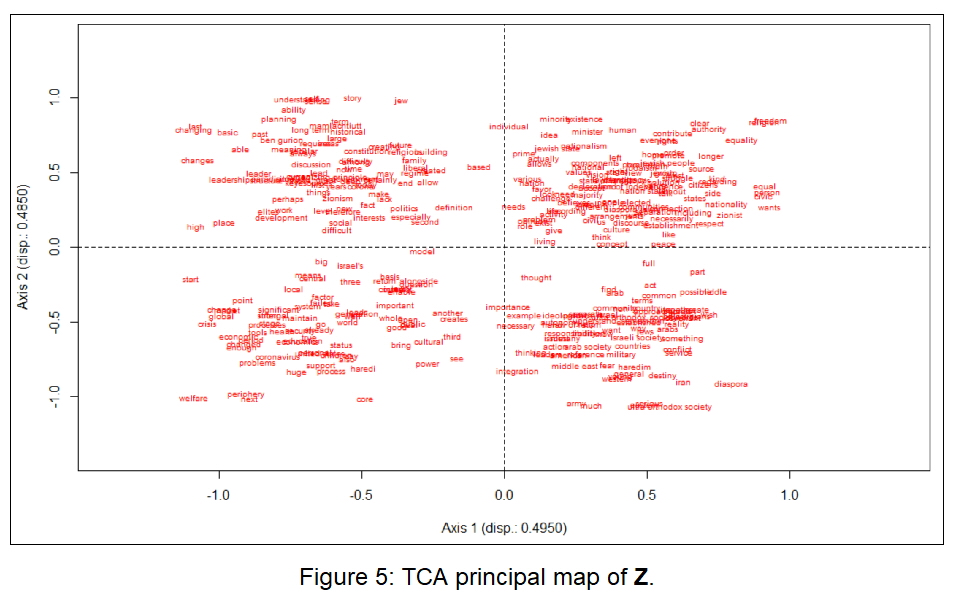}  
\end{figure}

\subsection{Taxicab Correspondence Analysis of \textbf{N or Z}}

Figure 5 displays the principal map of the 343 columns (words) obtained by
TCA of \textbf{N }or\textbf{\ Z}; where we see clearly 4 clusters. Some
elementary statistics of TCA of \textbf{N} are provided in Table 5. The QSR
values of axis 1 are quite low [(17.0,\textbf{\ }25.2), (-30.6,\ -38.8%
\textbf{), }25.5], which imply that these 4 clusters are quite heterogeneous
given the high sparsity value ($98.68\%$) of \textbf{N.}

In Figure 5, there are 4 major clusters of words in 4 quadrants, that we
designate by (sign1,sign2), where sign = Pos or Neg.

We name the subset of the contingency table \textbf{N} which corresponds to
the cluster in the upper right quadrant by \textbf{DataPosPos}. Similarly,
we name the 3 remaining subsets of \textbf{N} by \textbf{DataPosNeg}
(corresponding to th lower right cluster), and \textbf{DataNegPos} and 
\textbf{DataNegNeg}. In the following sections we study these 4 subsets.

\section{CA of each of the 4 data subsets}

Here we enumerate some important points of CA of the 4 datasets \textbf{%
DataPosPos}, \textbf{DataPosNeg}, \textbf{DataNegPos} and \textbf{DataNegNeg}%
.

a) The size and sparsity index of each of these 4 datasets are displayed in
Table 5, where we see that they are extremely sparse more than 96\%.

b) Similar to the CA map of \textbf{N }or\textbf{\ Z (}Figure 4), the CA map
(not shown) of each of these 4 datasets is difficult to interpret.

c) However, there is a \textbf{lateral ordering }in each of these 4 datasets,%
\textbf{\ }because the first singular values are close or larger than Benz%
\'{e}cri's empirical criterion of 0.835, as can be seen in Table 4: 0.812,
0.907, 0.866 and 0.862.

\section{TCA of each of the 4 data subsets}

TCA numerical results of the 4 datasets \textbf{DataPosPos}, \textbf{%
DataPosNeg}, \textbf{DataNegPos} and \textbf{DataNegNeg}, are summarized in
the Tables 5 through 9, and TCA maps are shown in Figures 6 through 9. Given
that the analysis of each of these 4 datasets are very similar, so we
present only the analysis of the dataset \textbf{DataPosPos}, then we
conclude.

\subsection{TCA of DataPosPos}

Figure 6 displays the principal map of the 106 words in the subset \textbf{%
DataPosPos}, where clearly we notice 4 clusters. Table 5 displays some TCA
statistics: We note that these QSR and dispersion $\delta _{\alpha }$\
values are uniformly higher in absolute value than the corresponding values
of TCA of \textbf{N }or\textbf{\ Z }shown in Table 5; which imply that the
subset \textbf{DataPosPos} is much more structural than the set \textbf{N }or%
\textbf{\ Z}. We designate the 4 subsets of DataPosPos by
DataPosPos(sign1,sign2), where sign = Pos or Neg. So we have 4 subsubsets
named:

DataPosPos(Pos ,Pos) = DataPPPP,

DataPosPos(Pos,Neg) = DataPPPN,

DataPosPos(Neg,Neg) = DataPPNN,

DataPosPos(Neg,Pos) = DataPPNP.

\begin{tabular}{lllll}
\multicolumn{5}{l}{\textbf{Table 5: TCA QSR (\%) values of the first 2
dimensions .}} \\ \hline\hline
\multicolumn{5}{l}{\textbf{(40+1) texts of size }$1605\times 343,\
sparsity=98.68\%$} \\ \hline
Axis $\alpha $ & $QSR_{\alpha }(V_{+}U_{+},V_{-}U_{-})$ & $QSR_{\alpha
}(V_{-}U_{+},V_{+}U_{-})$ & $QSR_{\alpha }$ & $\delta _{\alpha }$ \\ \hline
1 & (17.0,\textbf{\ }25.2) & (-30.6,\ -38.8\textbf{)} & 25.5 & \textbf{0.495}
\\ 
2 & (16.7,\ 19.5\textbf{)} & (-25.3\textbf{,\ }-27.1) & 21.4 & \textbf{0.485}
\\ \hline
\multicolumn{5}{l}{\textbf{DataPosPos of size }$431\times 106,$ $%
sparsity=97.13\%$} \\ \hline
Axis $\alpha $ & $QSR_{\alpha }(V_{+}U_{+},V_{-}U_{-})$ & $QSR_{\alpha
}(V_{-}U_{+},V_{+}U_{-})$ & $QSR_{\alpha }$ & $\delta _{\alpha }$ \\ \hline
1 & (22.2,\textbf{\ }30.2) & (-42.7,\ -55.5\textbf{)} & 33.4 & \textbf{0.624}
\\ 
2 & (25.5,\ 19.3\textbf{)} & (-31.6\textbf{,\ }-33.5) & 26.2 & \textbf{0.591}
\\ \hline
\multicolumn{5}{l}{\textbf{DataPosNeg of size }$418\times 78,$ $%
sparsity=96.78\%$} \\ \hline
Axis $\alpha $ & $QSR_{\alpha }(V_{+}U_{+},V_{-}U_{-})$ & $QSR_{\alpha
}(V_{-}U_{+},V_{+}U_{-})$ & $QSR_{\alpha }$ & $\delta _{\alpha }$ \\ \hline
1 & (26.3,\textbf{\ }28.8) & (-48.0,\ -60.6\textbf{)} & 36.3 & \textbf{0.676}
\\ 
2 & (19.4,\ 30.4\textbf{)} & (-32.8\textbf{,\ }-46.2) & 29.3 & \textbf{0.657}
\\ \hline
\multicolumn{5}{l}{\textbf{DataNegNeg of size }$383\times 77,$ $%
sparsity=97.13\%$} \\ \hline
Axis $\alpha $ & $QSR_{\alpha }(V_{+}U_{+},V_{-}U_{-})$ & $QSR_{\alpha
}(V_{-}U_{+},V_{+}U_{-})$ & $QSR_{\alpha }$ & $\delta _{\alpha }$ \\ \hline
1 & (32.7,\textbf{\ }23.3) & (-64.3,\ -46.0\textbf{)} & 36.1 & \textbf{0.671}
\\ \hline
2 & (19.8,\ 30.1\textbf{)} & (-30.9\textbf{,\ }-41.8) & 28.6 & \textbf{0.663}
\\ \hline
\multicolumn{5}{l}{\textbf{DataNegPos of size }$373\times 82,$ $%
sparsity=96.9\%$} \\ \hline
Axis $\alpha $ & $QSR_{\alpha }(V_{+}U_{+},V_{-}U_{-})$ & $QSR_{\alpha
}(V_{-}U_{+},V_{+}U_{-})$ & $QSR_{\alpha }$ & $\delta _{\alpha }$ \\ \hline
1 & (22.6,\textbf{\ }33.4) & (-45.5,\ -68.3\textbf{)} & 36.1 & \textbf{0.687}
\\ 
2 & (19.2,\ 31.6\textbf{)} & (-33.7\textbf{,\ }-51.9) & 30.2 & \textbf{0.653}
\\ \hline
\end{tabular}

\bigskip

\begin{figure}[h]
\label{fig:MagEngT}
\centering 
\includegraphics[scale=0.5]{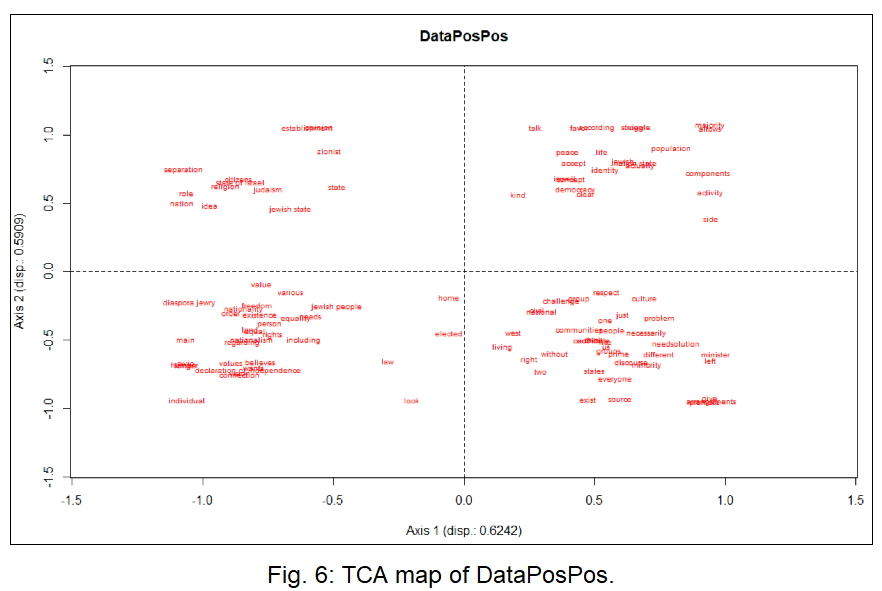}  
\end{figure}

The interpretation of the 4 clusters (as topics) in Figure 6 are outlined in
Table 6 by the words that contribute the most. To have a fuller
understanding of a topic composed of few words, one has to read the phrases
in which these words are found, as we applied in section 2 of DOI text
analysis. We explain this on the cluster DataPPPP, whose 23 columns (words
or terms) are plotted in the upper right quadrant of Figure 6.

\subsubsection{Columns that contribute the most}

Here are the 23 column labels ordered according to their contributions to
the principal plane (in parenthesis):

\bigskip 
\begin{tabular}{|c|}
\hline
activity (0.479); favor (0.480); kind (0.481); allows (0.639); \\ 
according (0.639); side (0.639); components(0.799); \\ 
nation-state (0.799); accept (0.800); struggle (0.959); \\ 
view (0.959); actually (0.959); talk (0.959); concept (0.960); \\ 
\textbf{life (1.12); peace (1.12); clear (1.12); majority (1.60);} \\ 
\textbf{population (1.76); identity (1.92); democracy (2.88);} \\ 
\textbf{israeli (3.04); jewish (7.03)} \\ \hline
\end{tabular}

\subsubsection{Rows that contribute the most}

Among the 106 phrases, here are the 6 rows (phrases) that contribute the
most to the principal plane (in parenthesis):

\textbf{R1020 (2.72): Interviewee 23, phrase 4:} A genuine \textbf{peace}
with open borders that will attract millions of people from all nations
creates a challenge and a strategic threat of being overtaken by others ---
a possible consequence of \textbf{peace}. And therefore, the lack of \textbf{%
peace} protects us from being flooded by foreigners and allows the \textbf{%
Jewish-Zionist-Israeli} state to prepare for \textbf{peace}.

\textbf{R489 (2.68): Interviewee 11, phrase 7:} From the Ultra-Orthodox
aspect,this is a deal that has only one side: if we were afraid to integrate
into the military, academia, work, in the \textbf{Israeli} sphere, so as not
to lose our Ultra-Orthodox \textbf{identity}, an opportunity was created
that we were willing to accept us.

\textbf{R906 (2.65): Interviewee 18, phrase 20: }The \textbf{majority} of
the Ultra-Orthodox community "lives in \textbf{peace}" with "\textbf{Jewish}
and \textbf{democratic}."

\textbf{R450+R871 (2.65): Interviewee 10(+17), phrase 20(+25): }Both \textbf{%
Jewish} and \textbf{democratic} (\textbf{Jewish- Democratic} was initially
seen as a contrast, the "hyphen" should be clarified.)

\textbf{R295 (2.04): Interviewee 8, phrase 12: }What has been clearly
forming over the past 100 years is the \textbf{Jewish-Israeli} group, which
has clear common \textbf{identity} characteristics, despite the various
groups within it (Ultra-Orthodox, etc.).

\textbf{R1700 (2.03): Interviewee 40, phrase 44: }We need to abandon our
zero-sum game struggle with Iran that does not lead to a positive, \textbf{%
clear} and practical solution, and find a different, creative approach of
regional understandings to learn perhaps from the Emirates who know how to
be at \textbf{peace} with both the Americans and Iran! A "both" solution!

\subsubsection{Interpretation}

We see that the core of this cluster concerns \textbf{Jewish-Israeli}
(10.07) society or state, and its characterization with the three words 
\textbf{democracy (}2.88\textbf{)}, \textbf{identity (}1.92\textbf{)} and 
\textbf{peace (}1.12\textbf{)}. Furthermore, \textbf{democracy} and \textbf{%
identity} pertain to the ultra-orthodox community; while \textbf{peace}
pertains within the society or with the neighboring or enemy states like
Iran.

This shows that the interpretation of a collection of words to represent a
topic depends on the context used in a collection of phrases
(\textquotedblright environment\textquotedblright ); this is the fundamental
structural concept of distributional analysis advanced by the linguist
Harris, whom Benz\'{e}cri (1966, subsection 4.2, page 336) cites
\textquotedblright The distribution of an element will be understood as the
sum of all its environment\textquotedblright\ and adds \textquotedblright
l'originalit\'{e} de Harris est d'op\'{e}rer syst\'{e}matiquement sur un
corpus (ou un texte unique) sans essayer de fabriquer des exemples selon les
besoins, ni de consid\'{e}rer le sens\textquotedblright ; that is
\textquotedblright The originality of Harris is to act on a given corpus or
a unique text\textquotedblright . Then Benz\'{e}cri compares his principle
of distributional equivalence on which he developed the mathematics of CA
and Harris's distributional analysis concept.

\begin{tabular}{c|cc}
\multicolumn{3}{c}{\textbf{Table 6: Summary of DataPosPos and its 4 clusters
(topics).}} \\ \hline
& size &  \\ \hline
\multicolumn{1}{|c|}{\textbf{DataPosPos}} & $431\times 106$ & $%
sparsity=97.13\%$ \\ \hline
\multicolumn{1}{|c|}{DataPPPP} & $108\times 23$ & $sparsity=91.63\%$ \\ 
\multicolumn{1}{|c|}{\textbf{TopicPPPP}} & \multicolumn{2}{|c}{\textbf{%
jewish,Israli,democracy,identity,peace,}} \\ \hline
\multicolumn{1}{|c|}{DataPPPN} & $125\times 38$ & $sparsity=94.23\%$ \\ 
\multicolumn{1}{|c|}{\textbf{TopicPPPN}} & \multicolumn{2}{|c}{\textbf{Prime
minister,promote,give,,contribute,discourse,}} \\ 
\multicolumn{1}{|c|}{} & \multicolumn{2}{|c}{\textbf{%
arrangements,solution,right,minority,two }} \\ \hline
\multicolumn{1}{|c|}{DataPPNN} & $105\times 32$ & $sparsity=93.57\%$ \\ 
\multicolumn{1}{|c|}{\textbf{TopicPPNN}} & \multicolumn{2}{|c}{\textbf{%
equality},\textbf{declaration of independence}} \\ \hline
\multicolumn{1}{|c|}{DataPPNP} & $93\times 13$ & $sparsity=86.85\%$ \\ 
\multicolumn{1}{|c|}{\textbf{TopicPPNP}} & \multicolumn{2}{|c}{\textbf{%
state-of-israel,separation,role,judaism,zionism}} \\ \hline
\end{tabular}

\bigskip

\begin{figure}[h]
\label{fig:MagEngT}
\centering 
\includegraphics[scale=0.5]{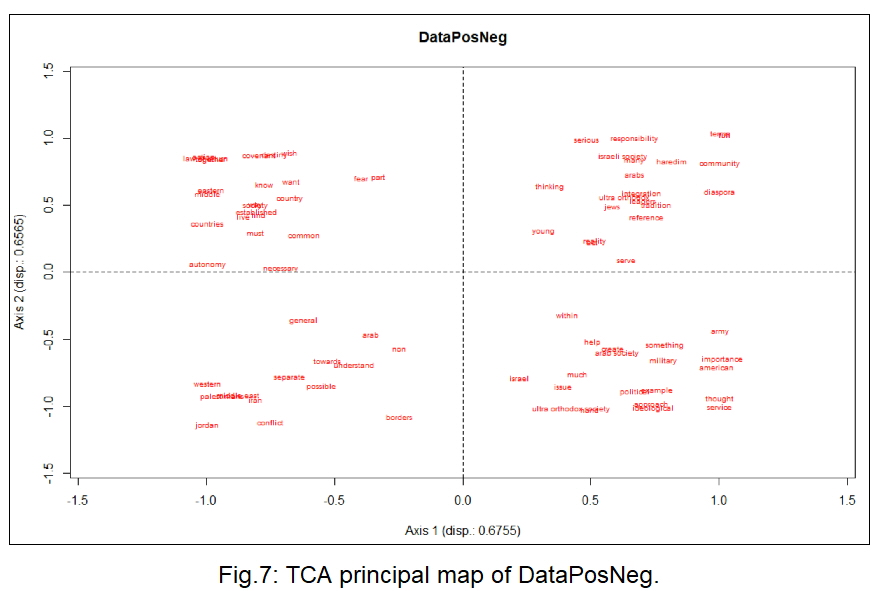}  
\end{figure}

\begin{tabular}{c|cc}
\multicolumn{3}{c}{\textbf{Table 7: Summary of DataPosNeg and its 4 clusters
(topics).}} \\ \hline
& size &  \\ \hline
\multicolumn{1}{|c|}{DataPosNeg} & $418\times 78$ & $sparsity=96.78\%$ \\ 
\hline
\multicolumn{1}{|c|}{DataPNPP} & $116\times 21$ & $sparsity=91.63\%$ \\ 
\multicolumn{1}{|c|}{\textbf{TopicPNPP}} & \multicolumn{2}{|c}{\textbf{%
ultra-orthodox,haredim,integration,israeli society,}} \\ 
& \multicolumn{2}{|c}{\textbf{community,jews,arabs}} \\ \hline
\multicolumn{1}{|c|}{DataPNPN} & $101\times 20$ & $sparsity=91.44\%$ \\ 
\multicolumn{1}{|c|}{\textbf{TopicPNPN}} & \multicolumn{2}{|c}{\textbf{%
Israel,military,army,service,issue,approach,}} \\ 
\multicolumn{1}{|c|}{} & \multicolumn{2}{|c}{\textbf{ultra-orthodox,arab}}
\\ \hline
\multicolumn{1}{|c|}{DataPNNN} & $95\times 14$ & $sparsity=89.47\%$ \\ 
\multicolumn{1}{|c|}{\textbf{TopicPNNN}} & \multicolumn{2}{|c}{\textbf{%
Palestinians,conflict,Middle-East,Iran,Arab}} \\ \hline
\multicolumn{1}{|c|}{DataPNNP} & $106\times 23$ & $sparsity=91.22\%$ \\ 
\multicolumn{1}{|c|}{\textbf{TopicPNNP}} & \multicolumn{2}{|c}{\textbf{%
society,countries,must,live,together,covenant,destiny}} \\ \hline
\end{tabular}

\bigskip

\begin{figure}[h]
\label{fig:MagEngT}
\centering 
\includegraphics[scale=0.5]{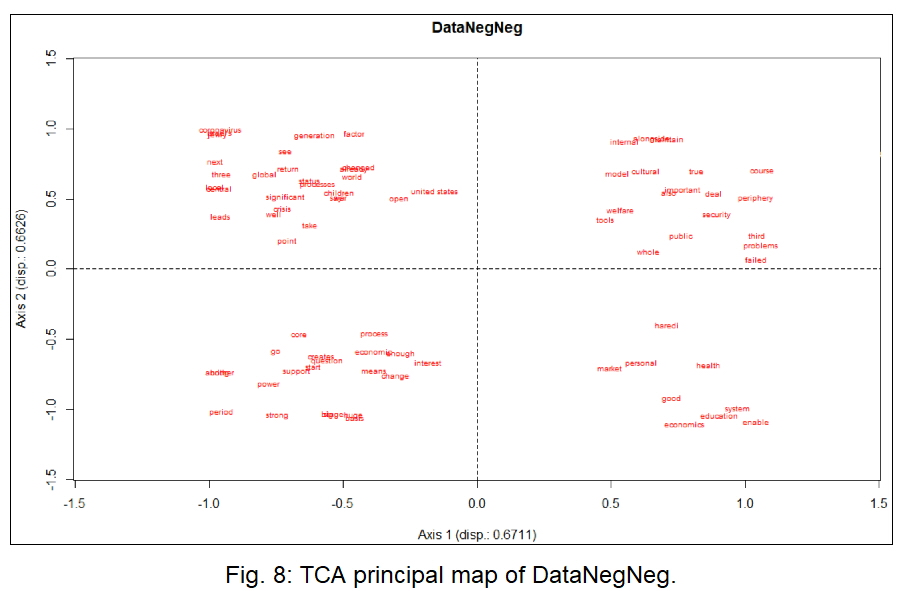}  
\end{figure}

\begin{tabular}{c|cc}
\multicolumn{3}{c}{\textbf{Table 8: Summary of DataNegNeg and its 4 clusters
(topics).}} \\ \hline
& size &  \\ \hline
\multicolumn{1}{|c|}{DataNegNeg} & $295\times 77$ & $sparsity=96.6\%$ \\ 
\hline
\multicolumn{1}{|c|}{DataNNPP} & $73\times 19$ & $sparsity=90.3\%$ \\ 
\multicolumn{1}{|c|}{\textbf{TopicNNPP}} & \multicolumn{2}{|c}{\textbf{%
important,security,deal,maintain,periphery,true,}} \\ 
& \multicolumn{2}{|c}{\textbf{public,cultural}} \\ \hline
\multicolumn{1}{|c|}{DataNNPN} & $65\times 9$ & $sparsity=84.1\%$ \\ 
\multicolumn{1}{|c|}{\textbf{TopicNNPN}} & \multicolumn{2}{|c}{\textbf{%
education,system,personal,health,economics,}} \\ 
\multicolumn{1}{|c|}{} & \multicolumn{2}{|c}{\textbf{market,enable}} \\ 
\hline
\multicolumn{1}{|c|}{DataNNNN} & $77\times 21$ & $sparsity=92\%$ \\ 
\multicolumn{1}{|c|}{\textbf{TopicNNNN}} & \multicolumn{2}{|c}{\textbf{%
economic,change,process,strong,power,basis}} \\ \hline
\multicolumn{1}{|c|}{DataNNNP} & $80\times 28$ & $sparsity=92.8\%$ \\ 
\multicolumn{1}{|c|}{\textbf{TopicNNNP}} & \multicolumn{2}{|c}{\textbf{%
crisis,world,coronavirus,global,processes}} \\ \hline
\end{tabular}

\begin{figure}[h]
\label{fig:MagEngT}
\centering 
\includegraphics[scale=0.5]{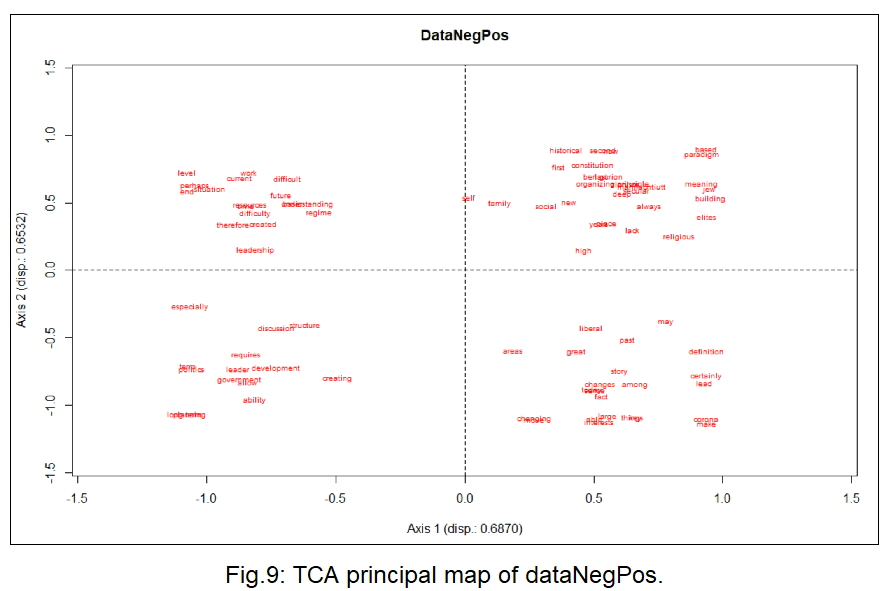}  
\end{figure}

\begin{tabular}{c|cc}
\multicolumn{3}{c}{\textbf{Table 9: Summary of DataNegPos and its 4 clusters
(topics).}} \\ \hline
& size &  \\ \hline
\multicolumn{1}{|c|}{DataNegPos} & $322\times 82$ & $sparsity=96.9\%$ \\ 
\hline
\multicolumn{1}{|c|}{DataNPPP} & $94\times 28$ & $sparsity=92.6\%$ \\ 
\multicolumn{1}{|c|}{\textbf{TopicNPPP}} & \multicolumn{2}{|c}{\textbf{%
zionism,organizing-principle,constitution,}} \\ 
& \multicolumn{2}{|c}{\textbf{religious,ben gurion,mamlachtiutt}} \\ \hline
\multicolumn{1}{|c|}{DataNPPN} & $80\times 23$ & $sparsity=92.5\%$ \\ 
\multicolumn{1}{|c|}{\textbf{TopicNPPN}} & \multicolumn{2}{|c}{\textbf{%
today,changes,large,past,story,act,sense,}} \\ 
\multicolumn{1}{|c|}{} & \multicolumn{2}{|c}{\textbf{%
make,able,lead,key,things,changing,}} \\ \hline
\multicolumn{1}{|c|}{DataNPNN} & $69\times 14$ & $sparsity=88.7\%$ \\ 
\multicolumn{1}{|c|}{\textbf{TopicNPNN}} & \multicolumn{2}{|c}{\textbf{\
ability,government,requires,longterm,planning}} \\ \hline
\multicolumn{1}{|c|}{DataNPNP} & $79\times 17$ & $sparsity=89.8\%$ \\ 
\multicolumn{1}{|c|}{\textbf{TopicNPNP}} & \multicolumn{2}{|c}{\textbf{%
current,situation,leadership,difficult,time,level}} \\ \hline
\end{tabular}

\subsection{Binary tree structure of the (40+1) texts}

Tables 6 through 9 show the existence of 16 clusters which we represent in a
complete binary tree with four levels, see Table 10, each level represents
the division of a subset of the data set by a principal dimension.

\begin{tabular}{cccc}
\multicolumn{4}{c}{\textbf{Table 10: Binary tree of TCA of (40+1) texts.}}
\\ \hline
level 4 & level 3 & level 2 & level 1 \\ \hline
PPPP & PPP & PP & P \\ 
PPPN &  &  &  \\ 
PPNP & PPN &  &  \\ 
PPNN &  &  &  \\ 
PNPP & PNP & PN &  \\ 
PNPN &  &  &  \\ 
PNNP & PNN &  &  \\ 
PNNN &  &  &  \\ 
NPPP & NPP & NP & N \\ 
NPPN &  &  &  \\ 
NPNP & NPN &  &  \\ 
NPNN &  &  &  \\ 
NNPP & NNP & NN &  \\ 
NNPN &  &  &  \\ 
NNNP & NNN &  &  \\ 
NNNN &  &  &  \\ \hline
\end{tabular}

\section{Adjusted sparsity index}

We attempt to shed further light on the following observation: How come CA
and TCA maps were interpretable of the DOI corpus, while the CA maps were
difficult to interpret (respectively the TCA maps were easily interpretable)
of the (40+1) corpus and its 4 subsets? Here we introduce the adjusted
sparsity index which may be helpful to understand this fact based on some
elementary basic results of Choulakian (2017) concerning CA and TCA of
sparse data tables. We remind the reader that the influence of rare
observations in CA is still an active area of research; see in particular
Nowak and Bar-Hen (2005), Greenacre (2013) and Lebart (2024).\bigskip

\textbf{Definition}: A contingency table \textbf{N} of size $I\times J$ with
only non-zero marginals is named sparsest if its sparsity in \% equals $%
100(1-1/min(I,J))$; and extremely sparse if its sparsity is very near to $%
100(1-1/min(I,J))$.\bigskip 

Note that in the above definition: First, the word sparsity means the real
sparsity of a data set computed after applying Benz\'{e}cri's principle of
distributional equivalence and not the apparent sparsity. Second, The
sparsest (I,J) tables are the tables of rank min(I,J) with exactly min(I,J)
non-zero entries. This means the sparsest tables are such that by permuting
the rows and the columns they become square diagonal data sets. Third, we
define for a given data set \textbf{N} 
\begin{equation*}
adjusted\ sparsity(\mathbf{N)}\text{ }in\text{ }\%=100\frac{sparsity\ of\ 
\mathbf{N}}{(1-1/min(I,J))}.
\end{equation*}

Table 11 provides a numerical summary of the 4 terms that appear in the
above definition concerning the 6 data sets which we analyzed in this paper
by CA and TCA; we also added the SACRED data set discussed by Choulakian and
Allard (2023). In Table 11, among the 7 data sets, only the CA map of the
first data set (DOI) was interpretable, while the CA maps of the last 6 data
sets were difficult to interpret. So based on Table 11, we conjecture the
following: Let \textbf{N} be a real data set that does not have an
underlying structure of blocks diagonal form, that is, the CA singular
values are strictly smaller than 1; then: \textit{The CA maps are difficult
to interpret if }$adjusted\ sparsity(\mathbf{N)\geq 98\%.}$

\bigskip 
\begin{tabular}{ccccc}
\multicolumn{5}{c}{\textbf{Table 11: The adjusted\ sparsity values of the 7
data sets.}} \\ \hline
data set & size & sparsity \% & upper boundary \% & adjusted\ sparsity(%
\textbf{N) }\% \\ \hline
DOI & $15\times 21$ & $88.25$ & $93.33$ & \textbf{94.55} \\ 
(40+1) & $1605\times 343$ & $98.68$ & $99.71$ & 98.9 \\ 
DataPosPos & $431\times 106$ & $97.13$ & $99.06$ & 98.06 \\ 
DataPosNeg & $418\times 78$ & $96.78$ & $98.72$ & 98.04 \\ 
DataNegNeg & $383\times 77$ & $97.13$ & $98.70$ & 98.4 \\ 
DataNegPos & $373\times 82$ & $96.9$ & $98.78$ & 98.10 \\ 
SACRED & $589\times \ 4864$ & $98.65$ & $99.83$ & 98.8 \\ \hline
\end{tabular}

\section{Conclusion}

In the early 1960's, Benz\'{e}cri developed CA specially for textual data
analysis; he considered the interpretability of the numerical outputs via
maps the most essential aspect of CA. We followed his approach in this
paper, where we focused on the interpretability of the numerical outputs via
maps and the interpretation of the anchor words (words that contribute the
most to the principal plane) in a cluster by choosing the phrases that
contribute the most to the creation of that cluster.

We summarize our results of this paper on the application of CA and TCA on
extremely sparse nonnegative data sets:

a) We presented the joint use of CA and TCA in the analysis-visualization of
2 data sets of different sizes: DOI text of small size (almost the size of
an ordinary poem) and (40+1) texts (almost the size of an ordinary
collection of poems).

b) TCA was able to reveal the underlying structure of extremely sparse
textual data sets as binary trees.

c) We attempted to shed further light on the interpretability of the CA maps
by the introduction of the adjusted sparsity index of a sparse data set.

Finally, we provide here an approach for integrating unstructured data from
different sources. This capability is becoming essential, for example, in
the analysis of textual data scraped from social media and transliterations
of text in call center recordings. In a long term perspective, the analysis
of big data requires an ability to integrate unstructured data with images,
voice and sensor based technologies. Here we handle nominal values captured
by a document term matrix.\bigskip

\bigskip \textit{Acknowledgement}

Choulakian's research has been supported by the Natural Sciences and
Engineering Research of Canada (Grant no. RGPIN-2017-05092).\bigskip

\textbf{References}

Allard J, Choulakian V (2019) \textit{Package TaxicabCA in R}

\ \ \ \ \ \ https://CRAN.R-project.org/package=TaxicabCA

Benz\'{e}cri JP (1966) Linguistique et math\'{e}matique. \textit{Revue
Philosophique de la France et de l'\'{E}tranger}, 156, pp. 309-374

Benz\'{e}cri JP (1981) \textit{Pratique de L'Analyse Des Donn\'{e}es: Volume
3, Linguistique \& Lexicologie}. Dunod

Choulakian V (2017) Taxicab correspondence analysis of sparse two-way
contingency tables. \textit{Italian Journal of Applied Statistics}, 29
(2-3), 153-179

Choulakian V, Kasparian S, Miyake M, Akama H, Makoshi N, Nakagawa M (2006) A
Statistical Analysis of the Synoptic Gospels. \textit{Journ\'{e}es
internationales d'Analyse statistique des Donn\'{e}es Textuelles}, 8: 281-88

Choulakian V, Allard J (2023) Visualization of extremely sparse contingency
table by taxicab correspondence analysis : A case study of textual data.
Available at https ://arxiv.org/pdf/2308.03079.pdf

Greenacre, M. (2013). The contributions of rare objects in correspondence
analysis. \textit{Ecology}, 94(1): 241-249

Greenacre M, Nenadic O, Friendly M (2022) \textit{Package ca in R}

https://CRAN.R-project.org/package=ca

Kenett RS, Gal R, Adres E, Ali N, Glickman H (2023) The Israeli society
common denominator: a text analytic study. \textit{Israel Affairs}, pp 1-11,
DOI: 10.1080/13537121.2023.2206250

Lebart L (2024) Low lexical frequencies in textual data analysis. In Beh,
Lombardo, Clavel (eds): \textit{Analysis of Categorical Data from Historical
Perspectives: Essays in Honour of Shizuhiko Nishisato}, pp 319-334, Springer.

https://doi.org/10.1007/978-981-99-5329-5\_19

Nowak E, Bar-Hen A (2005) Influence function and correspondence analysis. 
\textit{Journal of Statistical Planning and Inference}. 134: 26-35

Qi Q, Hessen DJ, Deoskar T, van der Heijden PGM (2023) A comparison of
latent semantic analysis and correspondence analysis of document-term
matrices. \textit{Natural Language Engineering}, pp 1--31,
doi:10.1017/S135132492300024\bigskip

\bigskip \textbf{Appendix}

\begin{figure}[h]
\label{fig:MagEngT}
\centering 
\includegraphics[scale=0.5]{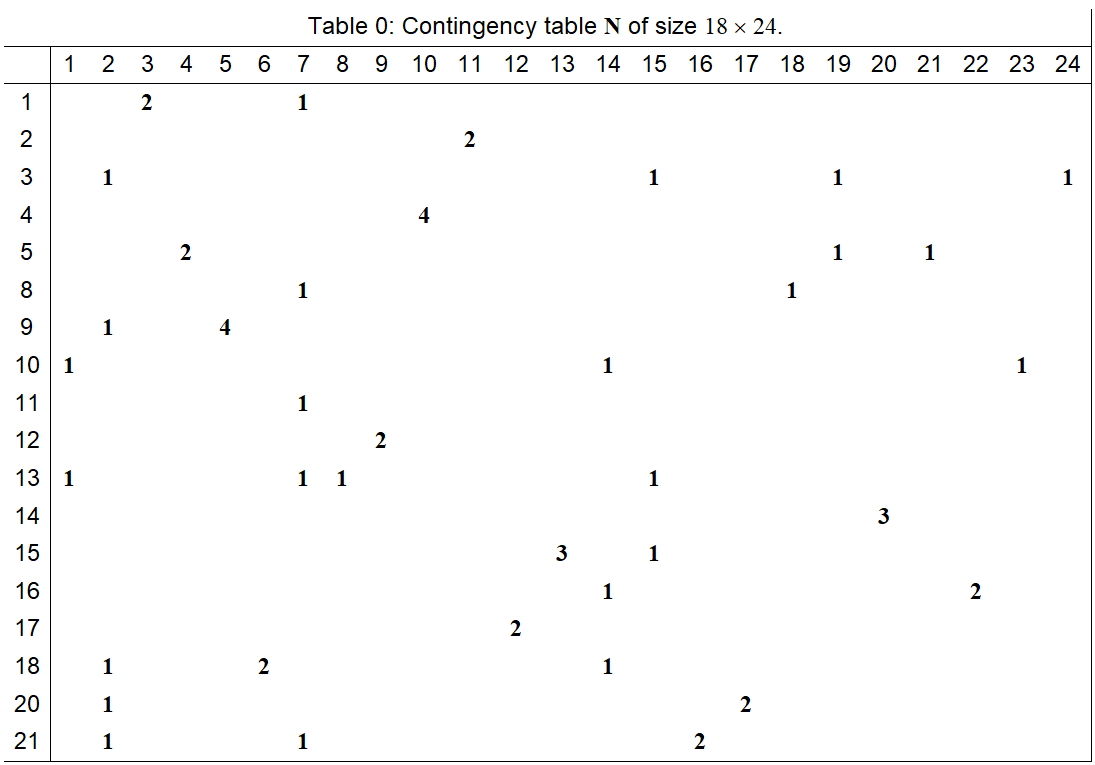}  
\end{figure}

\bigskip

\begin{figure}[h]
\label{fig:MagEngT}
\centering 
\includegraphics[scale=0.5]{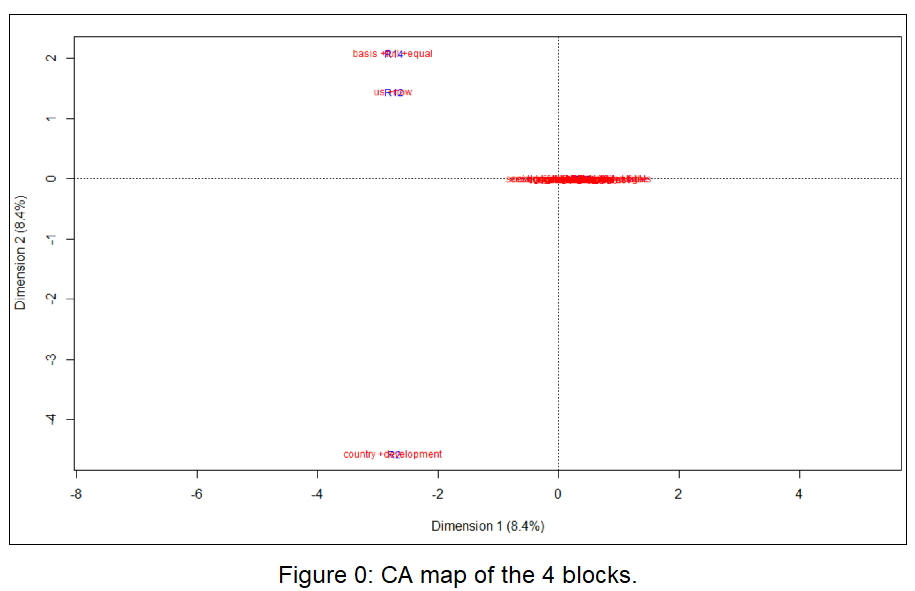}  
\end{figure}

\bigskip

\end{document}